\documentstyle[epsfig,preprint,aps,floats]{revtex}    
\newcommand{\mt}{m_{\tau}}

\newcommand{\gva}{\gamma_{VA}}
\newcommand{\spsi}{\sin\psi}
\newcommand{\cpsi}{\cos\psi}

\newcommand{\cpsiz}{\cos^{2}\psi}
\newcommand{\szpsi}{\sin 2\psi}

\newcommand{\sth}{\sin\beta}
\newcommand{\sthz}{\sin^{2}\beta}
\newcommand{\szth}{\sin 2\beta}

\newcommand{\cth}{\cos\beta}
\newcommand{\cthz}{\cos^{2}\beta}

\newcommand{\sph}{\sin\alpha}
\newcommand{\cph}{\cos\alpha}
\newcommand{\ke}{{K_{1}}} 
\newcommand{\kz}{{K_{2}}}
\newcommand{\kd}{{K_{3}}}
\newcommand{\kv}{{K_{4}}}
\newcommand{\kf}{{K_{5}}}

\newcommand{\keb}{{\overline{K}_{1}}}
\newcommand{\kzb}{{\overline{K}_{2}}}

\begin{document}
\draft
%
\preprint{
\font\fortssbx=cmssbx10 scaled \magstep2
\hbox to \hsize{
\hfill \vtop{   \hbox{\bf TTP96-43 }
                \hbox{\bf hep-ph/9609502\\}
                \hbox{September 1996}
                \hbox{}
                \hbox{              } }}
}
\title{CP Violation in Semileptonic $\tau$ Decays\\ with Unpolarized Beams }
\author{J.H.~ K\"uhn and E.~Mirkes}
\address{Institut f\"ur Theoretische Teilchenphysik, 
         Universit\"at Karlsruhe,\\ D-76128 Karlsruhe, Germany\\[2mm]}
\maketitle
\thispagestyle{empty}
\begin{abstract}
CP violating signals in semileptonic $\tau$ decays are studied.
These can be observed in decays of unpolarized single $\tau$'s
even if their rest frame cannot be reconstructed. No beam polarization 
is required. The importance of the two meson channel, in particular the 
$K\pi$ final state is emphasized.
\end{abstract}
%
\newpage
%

\section{Introduction}
CP violation has been experimentally observed only in the
$K$ meson system. The effect can be explained by a nontrivial complex
phase in the CKM flavour mixing matrix 
\cite{kmmat}. However, the fundamental origin of this CP violation
is still unknown.
In particular the CP properties of the third fermion
family are largely unexplored.
Production and decay of $\tau$ leptons might offer a particularly clean
laboratory to study these effects.
CP-odd correlations of the $\tau^-$ and $\tau^+$ decay products,
which originate from an electric dipole moment in the $\tau$ pair
production, are discussed in \cite{nachtmann,wermes}.
In this paper, we investigate the effects of possible
non-Kobayashi-Maskawa-type of CP violation, {\it i.e.} CP violation
effects beyond the Standard Model (SM) on semileptonic $\tau$ decays. 
Such effects could originate for
example from multi Higgs boson models \cite{mhiggs}.

CP violation effects in the $\tau\rightarrow 2\pi \nu$ decay mode
from this scalar sector 
have recently been discussed in terms of ``stage-two spin correlation
functions'' in \cite{nelson1} and in the case of polarized
electron-positron beams at $\tau$ charm factories in \cite{tsaicp}.
CP violation in the three pion channel has been also discussed in 
\cite{hagiwara} and in the $K\pi\pi$ and $KK\pi$ channels in \cite{koerner},
where the latter analysis is based on the ``$T-$odd'' correlations as
derived in \cite{km1} and the vector meson dominance parameterizations
in \cite{roger1}.

In the present paper we show that the structure function
formalism in \cite{km1} allows for a systematic analysis
of possible CP violation effects in the two and three
meson cases. Special emphasis is put on the $\Delta S=1$ transition
$\tau\rightarrow K\pi\nu_\tau$ where possible CP violating signals from
multi Higgs boson models would be signalled by a nonvanishing
difference between the structure functions
$W_{SF}[\tau^-\rightarrow (K\pi)^-\nu_\tau]$ and
$W_{SF}[\tau^+\rightarrow (K\pi)^+\nu_\tau]$.
Such a measurement is possible for unpolarized single $\tau$'s without
reconstruction of the $\tau$ rest frame and without
polarized incident $e^+e^-$ beams.
It is shown that this difference is proportional to
IM(hadronic phases) $\times$ IM(CP-violating phases),
where the hadronic phases arise from the interference of complex 
Breit-Wigner propagators, whereas the CP violating phases could 
arise from an exotic charged Higgs boson.
An additional independent test of CP violation
in the two meson case
would require the knowledge
of the full kinematics and  $\tau$
polarization. 

The paper is organized as follows: 
CP violating terms in the Hamiltonian for 
$\tau$ decays are discussed in section~II. It is shown that
CP violating signals induced through the exchange of an exotic intermediate
vector boson can only arise, if both vector and axial vector
hadronic currents contribute to the same final state,
{\it i.e.} for final state which are not eigenstates of $G$ parity and 
involve three or more mesons.
The kinematics and the relevant form factors and structure functions 
for tests of CP violation in the two meson case are presented
in section~III. CP violation effects in three meson final states
are briefly discussed in section~IV. 
Finally, 
it is shown in appendix A that the reconstruction of the full
kinematics is possible for the case, where one $\tau$
decays into one charged hadron and the second $\tau$ decays
into a $\rho$ or $K^*$ with the subsequent decay into a neutral and a charged
meson.

\section{CP Violation in $\tau$ Decays}
The Hamiltonian responsible for $\tau$ decays is decomposed into the
conventional term of the SM, 
denoted by $H_{SM}$, a CP violating term 
of similar structure, induced {\it e.g.} by the exchange of a vector boson, 
$H^{(1)}_{{CP}}$ and a CP violating term induced by scalar or pseudo
scalar exchange $H^{(0)}_{{CP}}$:
\begin{eqnarray}
H_{SM}&=&\cos\theta_c \frac{G}{\sqrt{2}}\,\,
[\bar{\nu}\,\,\gamma_\alpha(g_V-g_A\gamma_5)\,\,\tau]\,\,\,
\,\,\,[\bar{d}\,\,\gamma^\alpha(1-\gamma_5)\,\,u]
\hspace{10mm}+\hspace{11mm}h.c.\nonumber\\
H_{{CP}}^{(1)}&=&\cos\theta_c \frac{G}{\sqrt{2}}\,\,
[\bar{\nu}\,\,\gamma_\alpha(g_V^\prime-g_A^\prime\gamma_5)\,\,\tau]\,\,\,
\,\,\,[\bar{d}\,\,\gamma^\alpha(\chi_V^d-\chi_A^d\gamma_5)\,\,u]
\hspace{5mm}+\hspace{5mm}h.c.\label{hamilton}\\
H_{{CP}}^{(0)}&=&\cos\theta_c \frac{G}{\sqrt{2}}\,\,
[\bar{\nu}\,\,\,\,\,(g_S+g_P\gamma_5)\,\,\tau] \,\,\,\,\,\hspace{4mm}
[\bar{d}\,\,(\eta_S^d+\eta_P^d\gamma_5\,)\,u]\nonumber
\hspace{9mm}+\hspace{6mm}h.c.
\end{eqnarray}
plus a similar term with the complex parameters
$\eta^d,\chi^d$ replaced by $\eta^s,\chi^s$ for the
$\Delta S=1$ contribution.

The dominance of $(V-A)$ contributions to the leptonic current
in leptonic and semileptonic $\tau$-decays has been
demonstrated experimentally under fairly mild theoretical assumptions.
A pure $(V-A)$ structure of the leptonic current in
$H_{SM}$ will therefore be 
adopted for simplicity. A slight deviation from $(V-A)$ for the
hadronic current can in principle be masked by the form factors.
However, tight restrictions can be derived from the ratio
$\Gamma(\tau\rightarrow\nu\pi)/\Gamma(\tau\rightarrow\nu\pi\pi)$
using as input $f_\pi$ and the pion form factor from $e^+e^-$ 
annihilation\footnote{The relative sign
between hadronic vector and axial vector current is not fixed through this
consideration. It can be determined by interference measurements in the 
$K\pi\pi$ and $KK\pi$ channels, {\it e.g.} of the structure functions
$W_{F,G,H,I}$ \cite{km1}.}.
The study of CP violation will entirely
rely on interference terms between $H_{SM}$ and
$H_{{CP}}^{(1)},H_{{CP}}^{(0)}$. Interference
terms between the dominant $(V-A)$ leptonic current in $H_{SM}$
and a possible $V+A$ term in the leptonic current of
$H_{{CP}}^{(0,1)}$ are suppressed with the ratio between the 
mass of the $\tau$ neutrino and $m_\tau$. Therefore
only contributions from left--handed neutrinos will be included,
{\it i.e.} $g_V=g_A=g^\prime_V=g^\prime_A=g_S=-g_P=1$.

The hadronic matrix elements of the currents 
$\bar{d}\gamma^\alpha u$ (and similarly $\bar{d}\gamma^\alpha \gamma_5 u$)  
in $H_{SM}$ and 
$H_{{CP}}^{(1)}$ are of course identical. 
The spin zero part is closely related to the corresponding matrix element
of the scalar current in $H_{{CP}}^{(0)}$
through  the equation of motion:
\begin{equation}
Q^\alpha\,\,\bar{d}\gamma_\alpha u \,=\,(m_u-m_d)\,\bar{d}u;
\hspace{1cm}
-Q^\alpha\,\,\bar{d}\gamma_\alpha \gamma_5 u \,=\,
(m_u+m_d)\,\bar{d}\gamma_5u
\label{motion}
\end{equation}
with $Q^\alpha=i\partial^\alpha$.
The Hamiltonian in Eq.~(\ref{hamilton})
can thus be written in the form
\begin{eqnarray}
H &=&\cos\theta_c\frac{G}{\sqrt{2}}\,\,
\left[\bar{\nu}\,\gamma_\alpha(1-\gamma_5)\,\tau\right]\,\,\,
\left\{\left(
       (1+\chi_V^d)g^{\alpha\beta}+\frac{Q^\alpha Q^\beta}{m_\tau(m_u-m_d)}\,
       \eta_S^d\right)\,\,\,
       \bar{d}\,\gamma_\beta\, u\right. 
       \label{hamilt}\\
  &&\hspace{4.2cm} - \,\,\left.\left(
       (1+\chi_A^d)g^{\alpha\beta}+\frac{Q^\alpha Q^\beta}{m_\tau(m_u+m_d)}
       \eta_P^d\right)\,\,\,
       \bar{d}\,\gamma_\beta \gamma_5 \,u \right\}
       \hspace{5mm}+\hspace{5mm}h.c.
       \nonumber
\end{eqnarray}
and similarly for the Cabibbo suppressed mode.
From this form it is evident that $\chi$
and $\eta$ play a fairly different role.
Effects from $\chi_V$ and/or $\chi_A$ can only arise if both vector
and axial  hadronic currents contribute to the same final state
and hence only for final states which are not eigenstates of $G$
parity and involve three or more mesons.
Conversely,
for all two meson decays and, adopting isospin symmetry, even all multipion
states, CP violation cannot\footnote{In this aspect we disagree
with Ref.~\cite{tsaicp} where it has been claimed that CP 
violation in the $2\pi$  channel can be induced through 
the exchange of an exotic intermediate vector boson.}
arise from a complex $\chi$.
In contrast,  CP violation can arise from a complex
$\eta$, since $J=0$ and $J=1$ partial
waves are affected differently in this case.
For this reason contributions from nonvanishing $\chi$ will be ignored
in the following.

\section{Two Meson Decays: 
 Kinematics,  Form Factors {\protect{\newline}} and Structure Functions}
%
%
Transitions from the vacuum to two pseudoscalar mesons $h_1$ and
$h_2$ 
are induced through vector and scalar currents only, where the latter can 
be related to the former with the help of Eq.~(\ref{motion}).
Expanding this hadronic matrix element along the set of independent momenta
$(q_1-q_2)_\beta$ and $Q_\beta=(q_1+q_2)_\beta$
\begin{equation}
J_\beta=\langle h_1(q_1)h_2(q_2)|\bar{u}\gamma_\beta d|0 \rangle=
(q_1-q_2)^\delta\,T_{\delta\beta}\,\,F(Q^2) \,+\,
Q_\beta\, F_S(Q^2)
\label{jdef}
\end{equation}
the general amplitude for the strangeness conserving decay 
\begin{equation}
\tau^-(l,s) \rightarrow \nu(l^\prime,s^\prime) + h_1(q_1,m_1)+h_2(q_2,m_2)\>,
\label{twoh}
\end{equation}
can be written as\footnote{We suppress the superscript $d$ (or $s$)
in $\eta_S$ in the following.}
$$
{\cal{M}}=
  \cos\theta_{c}\frac{\,G}{\sqrt{2}}\,
  \bar{u}(l^\prime,s^\prime)\gamma_{\alpha}
           (1-\gamma_5)u(l,s)
  \left(g^{\alpha\beta}
  +\frac{Q^\alpha Q^\beta}{m_\tau(m_u-m_d)}\,\eta_S\right)
  \left[(q_1-q_2)^\delta\,T_{\delta\beta}\, F 
   + 
  Q_\beta\,F_S\right]
$$\mbox{}\vspace*{-8mm}
\begin{equation}
=
\cos\theta_{c}\frac{\,G}{\sqrt{2}}\,
  \bar{u}(l^\prime,s^\prime)\,\,\,\,\gamma_{\alpha}
           (1-\gamma_5)\,u(l,s) \,\,
  \left[\,\,(q_1-q_2)_\beta\,T^{\alpha\beta}\,\,\,  F   \, + \,
  Q^\alpha\,\,\,\tilde{F}_S\right]\hspace{2cm}
\label{me2h}
\end{equation}
with
\begin{equation}
\tilde{F}_S = \left(1+\frac{Q^2}{m_\tau(m_u-m_d)}\,\eta_S\right)\,\,F_S
\label{fs}
\end{equation}
and similarly for the $\Delta{S}=1$ part.
In Eq.~(\ref{me2h}) $s$ 
denotes the polarization 4-vector of the $\tau$ lepton
satisfying 
$
l_{\mu}s^{\mu}=0
$ and 
$
s_{\mu}s^{\mu}= -P^{2}.
$
$P$ denotes the polarization  of the $\tau$
in the $\tau$ rest frame with respect to its direction of flight 
in the laboratory frame and
$T^{\alpha\beta}$ is the  projector onto the spin one part
\begin{equation}
T^{\alpha\beta}=  g^{\alpha \beta} - \frac{Q^\alpha Q^\beta}{Q^2}  \>.
\end{equation}
As stated before, terms proportioal to $\chi$ do not contribute to CP
violation in the two meson case and have therefore been neglegted.


The representation of the hadronic amplitude
$\langle h_1h_2|\bar{u}\gamma_\beta d|0\rangle = (q_1-q_2)^\delta
\,T_{\delta\beta}\,\,  F   \, + \, Q_\beta\, \tilde{F}_S$
corresponds to a decomposition into spin one and spin
zero contributions,
{\it e.g.} the vector form factor $F(Q^2)$
corresponds to the $J^P = 1^-$ component
of the  weak charged current, and the scalar form factor 
$F_S(Q^2)$ to the $J^P = 0^+$ component. 
Up to the small isospin breaking terms, induced for example
by the small quark mass difference, CVC implies the
vanishing of $F_S$ for the two pion case. The small
$u$ and $d$ quark masses enter presumably the couplings from 
charged Higgs exchange and thus cancel the apparent enhancement
by the inverse power of $(m_u-m_d)$ in 
Eqs.~(\ref{hamilt},\ref{me2h},\ref{fs}).
The perspectives are more promising for the $\Delta{S}=1$ 
transition $\tau\rightarrow K\pi\nu$. The $J=1$ form factor $F$
is dominated by the $K^*(892)$ vector resonance contribution. 
However, in this case the scalar form factor $F_S$ is expected to
receive a sizable resonance contribution 
($\sim 5\%$ to the decay rate)
from the $K_0^*(1430)$ with $J^P=0^+$
\cite{kpi}.
In the subsequent discussion we will include both $\pi\pi$ and $K\pi$ 
final states. The corresponding $\tau^+$ decay is obtained from 
Eq.~(\ref{me2h})
through the substitutions 
\begin{equation}
(1-\gamma_5)\, \rightarrow \, (1+\gamma_5), \hspace{1cm}
  \eta_S      \, \rightarrow \, \eta_S^*.     \hspace{1cm}
\end{equation}

Reaction (\ref{twoh}) is most easily analyzed in the hadronic rest frame
$\vec{q}_{1}+\vec{q}_{2}=~0$.
After integration over the unobserved
neutrino direction,
the differential decay rate in the
rest frame of $h_{1}+h_{2}$ is given   by \cite{km1,kpi}
\begin{eqnarray}
d\Gamma(\tau^-\rightarrow 2h\nu_\tau)&=&
 \left\{
    \bar{L}_{B} {W}_{B} 
  + \bar{L}_{SA}{W}_{SA}
  + \bar{L}_{SF}{W}_{SF} 
  + \bar{L}_{SG}{W}_{SG}  \right\}\nonumber\\[3mm]
&&
\hspace{-1cm}
\frac{G^{2}}{2\mt} 
\bigl(^{\cos^2\theta_{c}}_{\sin^2\theta_{c}}\bigr) 
\frac{1}{(4\pi)^{3}}
\frac{(\mt^{2}-Q^{2})^{2}}{\mt^{2}}\,
\,|\vec{q}_{1}|\,\,
     \frac{dQ^{2}}{\sqrt{Q^{2}}}  \, \frac{d\cos\theta}{2}\,
     \frac{d\alpha}{2\pi}\,
     \frac{d\cos\beta}{2}\label{gamma}\>.
\end{eqnarray}
The coefficients $\bar{L}_X$ contain all  $\alpha, \beta$ and $\theta$
angular  and $\tau$-polarization dependence and will be specified
below.
The hadronic structure functions $W_X$,
$X\in\{B,SA,SF,SG\}$, depend only on
$Q^2$ and the form factors $F$ and $\tilde{F}_S$ of the hadronic current.
The dependence   can be obtained from
Eq.~(34) in \cite{km1} with the replacements
$
x_4  \rightarrow 2 \,\vec{q}_1 \>, 
F_3  \rightarrow -iF\, \>,
F_4  \rightarrow   \tilde{F}_S
$.
One has:
\begin{eqnarray}
W_B[\tau^-]    &=& 4 (\vec{q}_1)^2\,|F|^2 \\
W_{SA}[\tau^-] &=& Q^2\,  |\tilde{F}_S|^2  \label{wsa}         \\
W_{SF}[\tau^-]
 &=& 4\sqrt{Q^2}|\vec{q}_1|\, \mbox{Re}\left[F\tilde{F}_S^*\right]\\
W_{SG}[\tau^-]
 &=&-4\sqrt{Q^2}|\vec{q}_1|\, \mbox{Im}\left[F\tilde{F}_S^*\right]
\label{wsg}
\end{eqnarray}
where
$|\vec{q}_1|=q_1^z$ is the momentum of $h_1$ in
the rest frame of the hadronic system:
\begin{equation}
\vec{q}_{1}^{\,\,z}=\frac{1}{2\sqrt{Q^{2}}}\left(
[Q^{2}-m_1^{2}-m_2^{2}]^2-4m_{1}^{2}m_{2}^{2}\right)^{1/2}.
\end{equation}
The hadronic structure functions $W_X[\tau^+]$  are
obtained by the replacement $\eta_S\rightarrow\eta_S^*$
in $\tilde{F}_S$ in Eqs.~(\ref{wsa}-\ref{wsg},\ref{fs}).
CP conservation implies that all four structure functions are identical
for $\tau^+$ and $\tau^-$. With the ansatz 
for the form factors formulated in Eq.~(\ref{me2h})
CP violation can be present in $W_{SF}$ and $W_{SG}$
only and requires complex $\eta_S$.
As will be shown in the subsequent discussion
CP violation in $W_{SG}$ in maximal for fixed $\eta_S$ in the
absence of hadronic phases whereas $W_{SF}$ in contrast requires
complex $\eta_S$ and hadronic phases simultaneously.

We will now demonstrate that $W_{SF}$ can be measured in 
$e^+e^-$ annihilation experiments in the study of single unpolarized
$\tau$ decays even if the $\tau$ rest frame cannot be reconstructed.
In this respect the result differ from earlier studies of the two 
meson modes where either polarized beams and reconstruction
of the full kinematics \cite{tsaicp} or correlated fully reconstructed
$\tau^-$ and $\tau^+$ decays were required \cite{nelson1}.
The determination of $W_{SG}$, however, requires the knowledge of the full
$\tau$ kinematics and $\tau$  polarization.

For the definition and discussion of the angles and leptonic
coefficients $\bar{L}_X$ in Eq.~{(\ref{gamma})}
we therefore consider two different
situations, depending on whether the direction of flight of the $\tau$
in the hadronic rest frame [denoted by $\vec{n}_\tau$]
can be measured or not.\\[2mm]
\newpage
\noindent
\underline{ I) The $\tau$ direction in the hadronic rest frame is not known:}\\
In this case, the angle $\beta$ in Eq.~(\ref{gamma})
denotes the angle between the direction
of  $h_{1}\,$ ($\hat{q}_{1}=\vec{q}_{1}/|\vec{q}_{1}|$) 
and the direction of the laboratory $\vec{n}_{L}$ viewed
from the hadronic rest frame (see Fig.~1)
\begin{equation}
\cos\beta = \vec{n}_{L}\cdot \hat{q}_{1} \>.
\end{equation}
$\vec{n}_{L}$  is 
obtained from $\vec{n}_{L}=-\vec{n}_{Q}$, where
$\vec{n}_{Q}$ denotes the direction of the two-meson-system in the
laboratory.
The azimuthal angle $\alpha$ is not observable and has to be averaged out.

\setlength{\unitlength}{0.7mm}
\begin{figure}[t]               \vspace*{-6cm}
\begin{picture}(150,165)(-70,10)
\mbox{\epsfxsize8.0cm\epsffile[78 222 480 650]{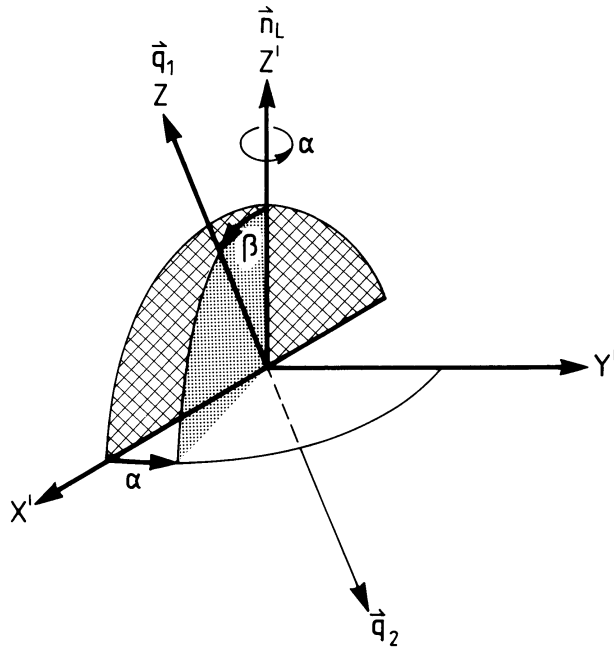}}
\end{picture}
\vspace*{5cm}
\caption{Definition of the angles $\alpha$ and $\beta$ in two meson
decays.
}
\label{fig1}
\end{figure}
The angle $\theta$  ($0\leq\theta\leq\pi$) 
in Eq.~(\ref{gamma})
which is only relevant for the analysis of polarized $\tau$'s,
is the angle between the
direction of flight of the $\tau$  
in the laboratory frame
and the direction of the  hadrons as seen in the $\tau$ rest frame.
The cosine of the angle $\theta$
is observable even in experiments, where
the $\tau$ 
direction cannot be measured experimentally.
This is because $\cos\theta$ 
can be  calculated \cite{KW84,km2,km1}
from the energy $E_{h}$ of the hadronic system
in the laboratory frame
\begin{equation}
\cos\theta = \frac{\left(2x\mt^{2}-\mt^{2}-Q^{2}\right)}{
          (\mt^{2}-Q^{2}) \sqrt{1-4\mt^{2}/s}}\>,
\label{cthdef}
\hspace{1cm}
\hspace{0.1cm}  x  = 2\frac{E_{h}}{\sqrt{s}} \>,
\hspace{1cm}    s = 4 E^{2}_{{beam}}\>.
 \end{equation}
Of particular importance  for the subsequent discussion is $\psi$, the angle 
 between the direction
of  the laboratory and the $\tau$ as seen from the hadronic rest frame.
Again the cosine of this angle can be
calculated from the hadronic energy $E_{h}$ \cite{KW84,km2,km1}.
One has:
  \begin{eqnarray}
\cos\psi &=&
       \frac{x(\mt^{2}+Q^{2})-2Q^{2}}{(\mt^{2}-Q^{2})\sqrt{x^{2}-4Q^{2}/s}} \>.
\label{cpsidef}
  \end{eqnarray}
The angular coefficients $\bar{L}_{B,SA,SF,SG}$ for $\tau^-$ decay
in Eq.~(\ref{gamma}) are given by \cite{km1}:
\begin{eqnarray}
\bar{L}_{B} &=&     {2}/{3}\,\ke\,+\,\kz
         \,-\,  2/3\,\keb\,(3\cthz-1)/2 \nonumber  \\
\bar{L}_{SA} &=&    \kz   \label{ldef}\\
\bar{L}_{SF}       &=&  - \kzb\,\cth \nonumber\\
\bar{L}_{SG}       &=&  0 \nonumber
\end{eqnarray}
with
\begin{eqnarray}
\ke &=&  1-\gva P \cos\theta -
        ({\mt^{2}}/{Q^{2}})\, (1+ \gva P \cos\theta)  \nonumber\\
\kz &=&  ({\mt^{2}}/{Q^{2}})\, (1+ \gva P \cos\theta)  \nonumber\\
\keb&=&  \ke \,(3\cpsiz-1)/2 - 3/2\,\kv\,\szpsi\nonumber \\
\kzb&=& \kz\,\cpsi\,+\,\kv\,\spsi \label{kalldef}\\[1mm]
\kv &=&   \sqrt{{\mt^{2}}/{Q^{2}}}\,\,\gva\, P \sin\theta \nonumber\\
\kf &=&  \sqrt{m_\tau^{2}/Q^{2}}\,\,P\sin\theta \nonumber
\end{eqnarray}
where
\begin{equation}
\gamma_{VA}= \frac{2g_{V}g_{A}}{g_{V}^{2}+g_{A}^{2}}
\label{gamvadef}
\end{equation}
($\kf$ in Eq.~(\ref{kalldef}) is needed in the later discussion).
In our case of purely left--handed leptonic currents $\gamma_{VA}=1$.
For $P=0$, the case relevant in the low energy
region $\sqrt{S}\approx 10$ GeV,  Eq.~(\ref{ldef}) simplifies to
\begin{eqnarray}
\bar{L}_{B}  &=&  \frac{1}{3}\left(2+\frac{\mt^2}{Q^2}\right)
            \, - \, \frac{1}{6}\left(1-\frac{\mt^2}{Q^2}\right)
                    \left(3\cpsiz-1\right)
                    \left(3\cos^2\beta-1\right)  \nonumber  \\
\bar{L}_{SA} &=&    \frac{\mt^2}{Q^2}    \\
\bar{L}_{SF} &=&    -\frac{\mt^2}{Q^2}\,\cpsi\,\cos\beta   \nonumber. 
\end{eqnarray}

Note that the angular coefficient $\bar{L}_{SG}$ vanishes, if the hadronic
rest frame is experimentally not known and only the distribution 
in $\beta$ is considered.

The differential rate (Eq.~(\ref{gamma})) for the
CP conjugated process
can be obtained from the previous results by
reversing all momenta $\vec{p}\rightarrow -\vec{p}$, 
the $\tau$ spin vector $\vec{s}\rightarrow -\vec{s}$,
the polarization  $P\rightarrow -P$, and the transition
$\gamma_{VA}\rightarrow-\gamma_{VA}$.
CP therefore  relates the differential decay rates
for $\tau^+$ and $\tau^-$ as follows:
\begin{equation}
d\Gamma[\tau^-](\gamma_{VA},P,W_X[\tau^-])
\rightarrow
d\Gamma[\tau^+](-\gamma_{VA},-P,W_X[\tau^+])
\end{equation}
Note that the coefficients $\bar{L}_X$ contain the full
$\gamma_{VA}$ and $P$ dependence.
From the  interference between the spin-zero spin-one terms
(denoted by the subscript $X=SF,SG$)
one can construct 
the following CP-violating quantities:
\begin{eqnarray}
\Delta_{X} &=& \frac{1}{2}
\left(
\bar{L}_{X}(\gamma_{VA},P)\, W_{X}[\tau^-]
-
\bar{L}_{X}(-\gamma_{VA},-P)\, W_{X}[\tau^+]
 \right)\\
&=&\bar{L}_{X}(\gamma_{VA},P)\,\frac{1}{2}
\left( W_{X}[\tau^-] -  W_{X}[\tau^+]\right)\\
&\equiv& \bar{L}_X\Delta W_X
\end{eqnarray}
As mentioned before the hadronic structure functions
$W_X[\tau^-]$ and $W_X[\tau^+]$ differ only in the complex parameter
$\eta_S$ in 
\begin{equation}
\tilde{F}_S[\tau^-]
 = \left(1+\frac{Q^2}{m_\tau(m_u-m_d)}\,\eta_S \right)\,\,F_S
\hspace{1cm}
\mbox{and}
\hspace{1cm}
\tilde{F}_S[\tau^+]
 = \left(1+\frac{Q^2}{m_\tau(m_u-m_d)}\,\eta_S^\ast \right)\,\,F_S
\end{equation}
and one obtains for the
only nonvanishing spin-zero spin-one term 
$L_{SF}W_{SF}$  
\begin{equation}
\Delta W_{SF}
\,\,= \,\,\, 4\sqrt{Q^2}|\vec{q}_1|\, \,
\frac{Q^2}{m_{\tau}(m_u-m_d)}\,\,\,\mbox{Im}\left(FF_S^*\right)\,\,
              \,     \mbox{Im}\left(\,\eta_S \right) \>.
\end{equation}
In essence this measurement analyses the difference in the
correlated energy distribution
of the mesons $h_1$ and $h_2$ from $\tau^+$ and $\tau^-$
decay in the laboratory. 
As already mentioned, $\Delta W_{SF}$ is observable for single
$\tau^+$ and $\tau^-$ decays without knowledge of the
$\tau$ rest frame. Any nonvanishing experimental result for 
$\Delta W_{SF}$ would be a clear signal of CP violation.
Note that a nonvanishing $\Delta W_{SF}$ requires nontrivial hadronic phases
(in addition to the CP violating phases $\eta_S$) in the form factors
$F$ and $F_S$. Such hadronic phases in $F$ ($F_S$) originate
in the $K\pi\nu_\tau$ decay mode from complex Breit Wigner
propagators for the $K^*$ ($K_0^*$) resonance. Sizable effects
of these hadronic phases are expected in this decay mode \cite{kpi}.

Once the $\tau$ rest frame is known and a preferred direction
of polarization exists
one may proceed further, determine also $W_{SG}$
and thus perform a second independent test for CP violation.
The subsequent analysis is performed for a $\tau$ with spin direction
$\vec{s}$.\\[2mm]
\noindent
\underline{ II) The $\tau$ direction in the hadronic rest frame is  known:}\\
As discussed in the appendix, 
the $\tau$ direction can be determined
in a large number of cases
with the help of vertex detectors.
In this case,
the vector $\vec{n}_L$ is replaced by $\vec{n}_\tau$ and
hence $\psi\Rightarrow 0$, whereas $\cos\theta=-\vec{s}\vec{n}_\tau$.
The angles $\alpha$ and $\beta$ are in this case  defined by
(see Fig.~1 with the replacement $\vec{n}_L\rightarrow\vec{n}_{\tau}$;
the $X^\prime-Z^\prime$--plane is defined by
the spin vector $\vec{s}$ and $\vec{n}_\tau$):
\begin{eqnarray}
\cos\beta&=&\vec{n}_{\tau}\cdot\hat{q}_{1}  \>,\\[2mm]
\cos\alpha&=&\frac{(\vec{n}_{\tau}\times\vec{s})\cdot
                   (\vec{n}_{\tau}\times\hat{q}_{1})}{
 |\vec{n}_{\tau}\times\vec{s}|\,\,|\vec{n}_{\tau}\times\hat{q}_1|}
                    \>,\label{calpha}\\[2mm]
\sin\alpha&=&-\frac{\vec{s}\cdot
                   (\vec{n}_{\tau}\times\hat{q}_{1})}{
|\vec{n}_{\tau}\times\vec{s}|\,\,|\vec{n}_{\tau}\times\hat{q}_{1}|}\>,
                          \label{salpha}
\end{eqnarray}
and the coefficients $\bar{L}_X$ for the $\tau^-$ decay
are given by:
  \begin{equation}
  \begin{array}{lcrl}
\bar{L}_{B} &=&     &\ke \sthz +  \kz - \kv \szth \cph        \\
\bar{L}_{SA}&=&     &\kz\\
\bar{L}_{SF}&=&  -   &\kz \cth- \kv \sth \cph\\
\bar{L}_{SG}&=&    &\kf \sth \sph 
  \end{array}
\label{ldef2}
  \end{equation}
The coefficients for $\ke,\kz,\kd,\kv,\kf$ for $\psi\neq 0$ are given in
(\ref{kalldef}).
For $P=0$, the case relevant in the low energy
region $\sqrt{S}\approx 10$ GeV,  Eq.~(\ref{ldef2}) simplifies to
  \begin{eqnarray}
\bar{L}_{B} &=& \left(1-\frac{m_\tau^2}{Q^2}\right)\,\sin^2\beta
                      +  \frac{m_\tau^2}{Q^2}   \\
\bar{L}_{SA}&=&  \frac{m_\tau^2}{Q^2}            \\
\bar{L}_{SF}&=&  -    \frac{m_\tau^2}{Q^2}\,\cos\beta \,      \\
\bar{L}_{SG}&=&     0 
  \end{eqnarray}
Evidently $W_{SG}$ cannot be measured in this case and thus no additional
test of CP violation is possible.
However, longitudinally polarized incident  beams \cite{tsaicp}
or the study of $\tau^+$ and $\tau^-$ correlations \cite{nelson1} could
allow to recover polarization.
The distributions in \cite{tsaicp} are in fact equivalent
to the product of $\bar{L}_{SG}W_{SG}$ if the angle $\alpha$ is defined
with respect to the $\tau$ spin vector through 
Eqs.~(\ref{calpha},\ref{salpha}).
Correlations between the $\tau^+$ and $\tau^-$
decay products may even allow to define the angle $\alpha$ without
$\tau$ polarization.

The differential rate (Eq.~(\ref{gamma})) for the
CP conjugated process
can be obtained from the previous results by
reversing all momenta $\vec{p}\rightarrow -\vec{p}$, 
the $\tau$ spin vector $ \vec{s}\rightarrow -\vec{s}$,
the polarization $P\rightarrow -P$, $\sin\alpha\rightarrow - \sin\alpha$
and the transition
$\gamma_{VA}\rightarrow-\gamma_{VA}$.
CP therefore  relates the differential decay rates
for $\tau^+$ and $\tau^-$ as:
\begin{equation}
d\Gamma[\tau^-](\sin\alpha,\gamma_{VA},P,W_X[\tau^-])
\rightarrow
d\Gamma[\tau^+](-\sin\alpha,-\gamma_{VA},-P,W_X[\tau^+])
\end{equation}
From the  interference between the spin-zero spin-one terms
one can construct 
the following CP-violating quantities:
\begin{eqnarray}
\Delta_{X} &=& \frac{1}{2}
\left(
\bar{L}_{X}(\sin\alpha,\gamma_{VA},P)\, W_{X}[\tau^-]
-
\bar{L}_{X}(-\sin\alpha,-\gamma_{VA},-P)\, W_{X}[\tau^+]
 \right)\\
&=& \bar{L}_{X}(\sin\alpha,\gamma_{VA},P)\,\frac{1}{2}
\left( W_{X}[\tau^-] -  W_{X}[\tau^+]\right)\\
&\equiv& \bar{L}_X\Delta W_X
\end{eqnarray}
\begin{equation}
\Delta W_{SF}
\,\,= \,\,\, 4\sqrt{Q^2}|\vec{q}_1|\,\,\,
   \frac{Q^2}{m_{\tau}(m_u-m_d)}\,\,\, \mbox{Im}\left(FF_S^*\right)
  \,\,    \mbox{Im}\left( \eta_S \right) \>.
\label{dwsf}
\end{equation}
\begin{equation}
\Delta W_{SG}
\,\,= \,\,\, 4\sqrt{Q^2}|\vec{q}_1|\,\,\,
\frac{Q^2}{m_{\tau}(m_u-m_d)}
\,\,\, \mbox{Re}\left(FF_S^*\right)
              \,\,     \mbox{Im}\left(\eta_S \right) \>.
\label{dwsg}
\end{equation}
Any observed nonzero value of these quantities would signal a 
true CP violation. Eqs.(\ref{dwsf}) and (\ref{dwsg}) show that the sensitivity
to CP violating effects in $\Delta W_{SF}$ and $\Delta W_{SG}$
can be fairly different depending on the hadronic phases.
Whereas $\Delta W_{SF}$ requires nontrivial hadronic phases
$\Delta W_{SG}$ is maximal for fixed $\eta_S$ in the absence of hadronic
phases.

\section{Three  Meson Decays}
The structure function formalism \cite{km1}
allow also for a systematic analysis of possible CP
violation effects in the three meson case.
Some of these effects have already been briefly discussed in
\cite{argonne}. The $K\pi\pi$ and $KK\pi$ decay modes
with nonvanishing vector \underline{and} axial vector
current are of particular importance for the detection of possible
CP violation originating from exotic intermediate vector bosons.
This would be signalled by a nonvanishing difference between
the structure functions $W_X(\tau^-)$ and $W_X(\tau^+)$
with $X\in\{F,G,H,I\}$. A difference in the structure functions
with $X\in\{SB,SC,SD,SE,SF,SG\}$ can again be induced through
a CP violating scalar exchange.
More details will be presented in a subsequent publication.

\appendix 
\section{Tau Kinematics in One Prong Decays
{\protect \\} from Impact Parameters}
The reconstruction of the full kinematics
is evidently a significant advantage
in  the analysis of $\tau$ decays . It has been shown in \cite{Kueimpact}
that this is possible for the $\tau^+(\rightarrow\pi^+\bar{\nu})
\tau^-(\rightarrow\pi^-\nu)$ final state 
if the tracks of $\pi^+$ and $\pi^-$
are experimentally measured---even if the production vertex
is not known. If the production vertex is known,
the knowledge of one track of $\pi^+$ or $\pi^-$
is sufficient. 
Also three prong decays with measured tracks allow the full kinematic
reconstruction.
In this appendix the discussion of \cite{Kueimpact}
is generalized to the case, where one $\tau$
decays into one charged hadron
and the second $\tau$ decays into $\rho$ or $K^\ast$ with the subsequent
decay into a neutral and a charged meson. 
For definiteness assume the $\tau^-$ decays into 
$\pi^-\nu$ and the $\tau^+$  decays into
$\rho^+(\rightarrow \pi^+\pi^0) \bar{\nu}$.
This is a generalization of the situation discussed in \cite{Kueimpact},
where both $\tau$'s were assumed to decay into one charged hadron each.
In fact, the method is also applicable in the multibody
one prong decay, if the momenta of all neutral decays are 
determined experimentally.

Assume that the momenta of all hadrons are measured and, furthermore, 
the tracks of $\pi^+$ and $\pi^-$ are also measured.
We shall demonstrate that also in this case the impact vector allows
to reconstruct the original direction of $\tau^+$ and $\tau^-$
in most of the cases.

We use the following notation:
$\vec{n}_+$ and  $\vec{n}_-$  denote unit vectors into
the directions of the $\rho^+$ and the $\pi^-$ momentum.
$\vec{r}$ denotes the direction of the $\pi^+$ track.
The cosines of the angles $\theta_\pm^L$ between the $\tau^\pm$ directions
and the hadronic systems in the lab frame are denoted by
$c_\pm\equiv \cos\theta_\pm^L=\vec{n}_{\tau_\pm}\vec{n}_{\pm}$.
$c_+$ and $c_-$ can be calculated from the energy 
and mass of the hadronic system as given in Eq.~(1) of \cite{Kueimpact}.

In a first step, we determine the vector  $\vec{d}$ which connects
the $\tau_+$ and
$\tau_-$ decay vertex.
As shown in \cite{Kueimpact}
$\vec{d}$ can be geometrically be reconstructed from $c_+, c_-, 
\vec{n}_+, \vec{n}_-$ up to  a two-fold ambiguity.
Only the directions, but not the length of $\vec{d}$ can be 
obtained in this way, therefore we define $\vec{D}=\vec{d}/|\vec{d}|$.
From Eq.~(9) in \cite{Kueimpact} we find the unit vector
\begin{eqnarray}
\vec{D}&=& \vec{D}_{min} + \frac{1}{1-(\vec{n}_+\vec{n}_-)^2}
\left[ -\left(c_+(\vec{n}_+\vec{n}_-)+c_-\right)\,\,\vec{n}_-
       +\left(c_-(\vec{n}_+\vec{n}_-)+c_+\right)\,\,\vec{n}_+\right]\\
       &\equiv&
       \vec{D}_{min} + \vec{D}_{0}
       \label{ddef}
\end{eqnarray}
$\vec{D}_{min}$ denotes the direction of the vector which connects 
the points of closest approach between the straight lines through the decay
vertex of $\tau^+$ and $\tau^-$ directions 
$\vec{n}_+$ and $\vec{n}_-$, respectively
and hence is normal to 
$\vec{n}_+$ and $\vec{n}_-$.
$\vec{D}_0$ is the remaining piece of $\vec{D}$
which by construction lies in the plane spanned by $\vec{n}_+$
and $\vec{n}_-$.

Note that  $ \vec{D}_{min} \perp \vec{D}_{0}$ by construction
and hence
\begin{eqnarray}
\vec{D}_{min}&=& 
      \pm \frac{(\vec{n}_+\times\vec{n}_-)}{
                |\vec{n}_+\times\vec{n}_-|}\,\,|\vec{D}_{min}|
\end{eqnarray}
From Eq.~(\ref{ddef}) and $\vec{D}^2=1, \, \vec{D}_{min}\cdot\vec{D}_0=0$
one has
\begin{eqnarray}
\vec{D}_{min}^2&=& 
       1-\vec{D}_0^2\\
               &=&
       \frac{1}{1-(\vec{n}_+\vec{n}_-)^2}
       \left\{1-\left[ c_+^2+c_-^2 + (\vec{n}_+\vec{n}_-)^2
       + 2c_+c_- (\vec{n}_+\vec{n}_-)\right]   \right\}
\end{eqnarray}
With
$|\vec{n}_+\times\vec{n}_-|=\sqrt{1-(\vec{n}_+\vec{n}_-)^2}$
on obtains
\begin{eqnarray}
\vec{D}&=&\pm\,\,
       \frac{ (\vec{n}_+\times\vec{n}_-)}{(1-(\vec{n}_+\vec{n}_-)^2)}
\sqrt{1-\left[ c_+^2+c_-^2 + (\vec{n}_+\vec{n}_-)^2
       + 2c_+c_- (\vec{n}_+\vec{n}_-)\right] } \nonumber\\
       & &\,\,\,-\,
       \frac{1}{ (1-(\vec{n}_+\vec{n}_-)^2)}
       \left[ \left(c_+(\vec{n}_+\vec{n}_-)+c_-\right)\,\,\vec{n}_-
       -\left(c_-(\vec{n}_+\vec{n}_-)+c_+\right)\,\,\vec{n}_+\right] \\
    &=& \pm\vec{D}_\perp + \vec{D}_0
\end{eqnarray}
No information on the tracks has been used.
In a second step one predicts from measured momenta
the vector $\vec{v}$ of minimal
distance between the positive and negative track with the help of 
$\vec{r}, \vec{n}_-,$ and $\vec{D}$.
Depending on the two choices for $\vec{D}$ one will obtain two predictions 
for the direction of $\vec{v}$. Frequently, 
only one of them will be compatible 
with the actual measurement. 

In order to calculate the vector $\vec{v}$ we choose a coordinate frame
where the $\tau_-$ decay point lies at the origin and the
negative and positive tracks are parameterized as
$\lambda \vec{n}_-$ and $\vec{D}+\lambda\vec{r}$ respectively.
[In reality one should scale $\vec{D}$ with $|\vec{d}|$, the 
actual decay length of $\tau^+$ plus $\tau^-$. However, since we
are only interested in directions, this factor drops out.]
A straightforward geometrical calculation predicts
\begin{equation}
\vec{v}=\vec{D}+\left[
\left( (\vec{D}\vec{r})\,(\vec{r}\vec{n}_-)
      -(\vec{D}\vec{n}_-)\right)\,\vec{n}_- 
      \,+\,
      \left((\vec{D}\vec{n}_-)\,(\vec{n}_-\vec{r})
      -(\vec{D}\vec{r})\right)\,\vec{r}\right]
      \frac{1}{1-(\vec{r}\vec{n}_-)^2}
      \label{vdef}
\end{equation}
Up to now no use has been made of track measurements.
At this point $\vec{r}$ is given by the momentum of the $\pi^+$.
Inserting the two solutions $\vec{D}_\pm=\pm\vec{D}_\perp+\vec{D}_0$
one obtains two predictions for $\vec{v}$.
\begin{equation}
\vec{v}_1=\vec{D}_+ +\left[
\left((\vec{D}_+\vec{r})\,(\vec{r}\vec{n}_-)
      -(\vec{D}_+\vec{n}_-)\right)\,\vec{n}_- 
      \,+\,
      \left((\vec{D}_+\vec{n}_-)\,(\vec{n}_-\vec{r})
      -(\vec{D}_+\vec{r})\right)\,\vec{r}\right]
      \frac{1}{1-(\vec{r}\vec{n}_-)^2}
      \label{v1def}
\end{equation}
\begin{equation}
\vec{v}_2=\vec{D}_- +\left[
\left((\vec{D}_-\vec{r})\,(\vec{r}\vec{n}_-)
      -(\vec{D}_-\vec{n}_-)\right)\,\vec{n}_- 
      \,+\,
      \left((\vec{D}_-\vec{n}_-)\,(\vec{n}_-\vec{r})
      -(\vec{D}_-\vec{r})\right)\,\vec{r}\right]
      \frac{1}{1-(\vec{r}\vec{n}_-)^2}
      \label{v2def}
\end{equation}
Note that both solutions $\vec{v}_1$ and $\vec{v}_2$ are normal to $\vec{n}_-$
and parallel or antiparallel to each other, 
i.e. $\vec{v}_1\times\vec{v}_2=0$.

The next step is to compare the measured $\vec{v}$ with the calculated
solutions $\vec{v}_1, \vec{v}_2$ and select the proper one.
In order that this strategy can work we have to restrict ourselves to 
the case, where  the two results for $\vec{v}$ in
Eqs.~(\ref{v1def},\ref{v2def})
correspond  to two opposite directions for $\vec{v}$ 
(and not only to different lenghts which would not allow to 
resolve the ambiguity).

Using Monte Carlo methods
we have checked for the $\pi^-\pi^0$ configuration that 
this is possible for
100 \%  of the events with $\pi^-\pi^0$ at threshold,
about 50\% for events with $m_{\pi\pi}\sim m_{\rho}$ 
(the dominant configuration!) and zero at the kinematical
endpoint $m_{\pi\pi}=m_{\tau}$, as illustrated in Fig.~2.
\setlength{\unitlength}{0.7mm}
\begin{figure}[t]               \vspace*{-2cm}
\begin{picture}(150,165)(-70,1)
\mbox{\epsfxsize10.0cm\epsffile[78 222 480 650]{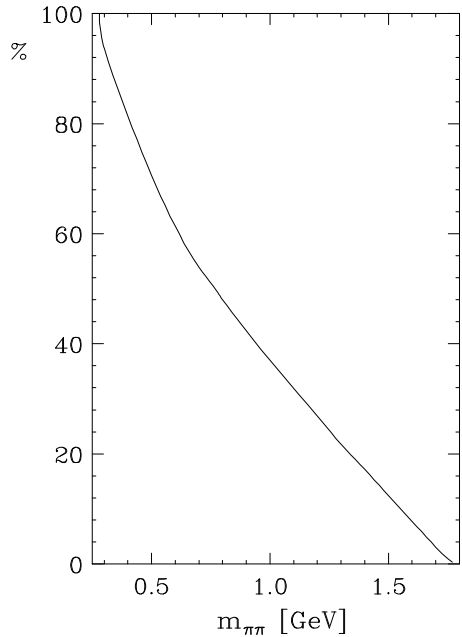}}
\end{picture}
\caption{
Percentage of events where the difference in the direction of 
$\vec{v}_1$ and $\vec{v}_2$
allows the full reconstruction.
}
\label{fig2}
\end{figure}
On the other hand, in the limit $m_{\pi\pi}\rightarrow m_{\tau}$
the difference between the two solutions per se shrinks to zero, such that
an approximate reconstruction seems possible in this case again.\\[12mm]
{\bf Acknowledgements}\\
The work of E.~M. was supported in part  
by DFG Contract Ku 502/5-1.
%

%

\end{document}